\documentclass[aps,prc,twocolumn,showpacs,showkeys,nofootinbib,superscriptaddress]{revtex4-1}

\usepackage{longtable}                  
\usepackage{graphicx}
\usepackage{amsmath,amsfonts,amssymb}

\usepackage{natbib} 
\usepackage{setspace}
\usepackage{xspace}
\usepackage{cancel}
\usepackage{physics}
\usepackage{tensor,braket,color,bm,slashed}

\usepackage{txfonts}
\usepackage{mathrsfs}

\usepackage[utf8]{inputenc}

\begin{document}

\title{
 \begin{flushright}
  \rightline{\normalsize APCTP-18-9/4, LFTC-18-11/32} 
 \end{flushright}
Electroweak properties of pions in a nuclear medium}

\author{Parada T. P. Hutauruk}
\email{parada.hutauruk@apctp.org}
\affiliation{Asia Pacific Center for Theoretical Physics, Pohang, Gyeongbuk 37673, Korea }

\author{Yongseok Oh}
\email{yohphy@knu.ac.kr}
\affiliation{Department of Physics, Kyungpook National University, Daegu 41566, Korea}
\affiliation{Asia Pacific Center for Theoretical Physics, Pohang, Gyeongbuk 37673, Korea }

\author{K. Tsushima}
\email{kazuo.tsushima@gmail.com}
\affiliation{Laborat\'orio de F\'isica Te\'orica e Computacional, Universidade Cruzeiro 
do Sul, 01506-000, S\~ao Paolo, SP, Brazil }
\affiliation{Asia Pacific Center for Theoretical Physics, Pohang, Gyeongbuk 37673, Korea }


\begin{abstract}

The charge form factor and weak decay constant of the pion as well as the pion-quark coupling constant
in symmetric nuclear matter are explored in the framework of the Nambu--Jona-Lasinio model,
where the pion is described as a bound state of dressed quark-antiquark pair 
obtained by the Bethe-Salpeter equation. 
For the in-medium current quark properties, we adopt the quark-meson coupling model, 
which describes successfully many hadron properties in a nuclear medium. 
The pion decay constant and the pion-quark coupling constant are found to decrease 
with increasing density as well as the magnitude of the light quark condensate.
But the pion mass is found to be insensitive to density up to $1.25$ times the normal nuclear density.
The pion charge form factor in the space-like region is also explored and is found to have 
a similar $Q^2$ dependence as the form factor in vacuum showing $1/Q^2$-behavior in large 
$Q^2$ region, where $Q^2$ is the negative of the four-momentum transfer squared.
The modifications of the charge radius of the charged pion in nuclear matter are then estimated and the
root-mean-square radius at the normal nuclear density is predicted to be larger 
than that in vacuum by about 20\%.  

\end{abstract}

\maketitle

\section{Introduction} 
\label{intro}

Pions, the Goldstone bosons emerged as a consequence of spontaneously breaking of global chiral symmetry 
in the favor SU(2) sector, play special roles in understanding the strong interactions, 
i.e., Quantum Chromodynamics (QCD),
at low energies~\cite{Ioffe05,Goldstone61,GSW62b,LP71b}. 
In particular, the pion electromagnetic form factor would provide us with information  
on the non-perturbative aspects of the pion internal structure as well as on the underlying quark-gluon 
dynamics~\cite{HBCT18,BR05}. 
There have been many studies devoted to understand the internal structure of the pion 
in free space based on various types of 
phenomenological approaches~\cite{Zovko74,FJ79,BM88a,BT02,NK07b,DDEF12,
CCRST13,NBC14,CBT14,HCT16,QCMR17,HBCT18,Hutauruk18}
and in lattice QCD simulations~\cite{HKLLWZ07,OKLMM15,HPQCD-17,PACS-17}.
Experimentally, there have been many data accumulated for the pion form 
factor~\cite{ABBB84,NA7-86,JLab_Fpi-00,JLab_Fpi-06,JLAB_Fpi-08} 
and new experimental measurements have been proposed~\cite{Horn18}.

However, all those studies focused on the pion form factor in vacuum and only a few studies have been 
reported so far on the pion form factors in nuclear medium.
We note that, in Ref.~\cite{DTERF14}, the authors including one of us used a light-front constituent 
quark model to describe the pion in vacuum as well as in medium. 
Although it gave us some insights on the properties of in-medium pions, there are a few aspects which 
require further investigations.
Namely,
(i) in the light-front constituent quark model, the constituent quark mass value in vacuum is an input 
and treated as a free parameter 
and 
(ii) the vacuum in the light-front approach is ``trivial'' and there is no quark condensate, which cannot make 
a direct relation with spontaneous breaking of chiral symmetry of the vacuum.
Therefore, although the study of Ref.~\cite{DTERF14} gave a first step in studying the pion properties 
in medium, we need to improve and test it further by making connections with the spontaneous breaking of chiral 
symmetry, which should be related to the basic features of the strong interactions.
In the present article, we address this point and, for this purpose, we make use of the Nambu--Jona-Lasinio 
(NJL) model as an effective theory of QCD, which describes the spontaneous breaking of chiral symmetry and 
offers the dynamically generated constituent quark mass based on quark condensates.

Some observations, such as the EMC effect~\cite{EMC83b}, indicate that the internal structure of hadrons 
may change in nuclear medium and also imply partial restoration of chiral symmetry in nuclear medium~\cite{JHK08}. 
The phenomenon of medium modifications is, therefore, one of most interesting subjects in nuclear and 
hadron physics~\cite{BR91,STT05,HH08,LMM09,MNB17}. 
In particular, how the properties of hadrons and their internal structures change by the surrounding 
nuclear environment is one of the important issues, which collects special 
attentions~\cite{BR91,HH08,LMM09,TSTW97,STT05,MNB17} in connection with the partial restoration of 
chiral symmetry in a strongly interacting environment~\cite{BR96,JHK00b,KY03,HR16}.  
The order parameters of this phenomena are the light-quark condensates in nuclear medium~\cite{JHK07}
and their changes in nuclear medium are main driving forces for the change of the hadron properties in nuclear 
medium~\cite{CRG93}. 
However, finding the clear experimental evidence of partial restoration of chiral symmetry is still challenging.

Spontaneous breaking of global and local chiral symmetry are important in understanding QCD and the 
former generates the pseudoscalar meson nonet of massless Goldstone bosons. 
The explicit breaking of the U(1) axial symmetry selectively shifts up the $\eta_0^{}$-singlet mass
leaving the SU(3) flavor octet of pions, kaons, and $\eta_8^{}$ to be massless. 
Dynamical symmetry breaking in QCD further produces most of the nucleon mass with the explicit 
chiral symmetry breaking. 
Then, the flavor symmetry breaking leads to the experimentally observed patterns in low-lying hadron spectra~\cite{BM88,KLVW90}. 
Since chiral symmetry has a big impact, in particular, on the low-lying hadron mass spectrum, the partial restoration of 
chiral symmetry 
in a strongly interacting medium is important to understand the change of hadron properties in nuclear 
medium.

As pions are the lightest bound states of dressed quark-antiquark pairs, we first focus on the electroweak 
properties of pions in nuclear medium in this exploratory study by calculating the space-like electromagnetic 
form factor and weak decay constant of the in-medium pion as well as the pion-quark coupling constant 
in symmetric nuclear matter.
For this purpose, we use the NJL model that is one of the widely used chiral effective quark models of 
QCD with numerous successes in studying meson  
properties~\cite{BM88,BJM88,KLVW90,TTKK90,Klevansky92,HCT16,HBCT18}. 
Recently, the NJL model has been successfully used to study the pion form factor in 
vacuum~\cite{NBC14,CBT14,HCT16}. 
In the present work, following the formalism of Ref.~\cite{HCT16}, we study the electroweak properties of pions in 
symmetric nuclear matter.
For the in-medium modified light quark properties that are necessary to study in-medium pions,  
we make use of the quark-meson coupling (QMC) model~\cite{GST18,SGRT16}. 
The in-medium quark properties obtained in this model are used as inputs to study the properties of in-medium pions
in the NJL model by solving the gap equation for the dynamical quark mass and the Bethe-Salpeter equation for the pion bound state
in nuclear medium.

This paper is organized as follows. 
We begin with a short review on the formalism to compute pion properties in the NJL model in 
Sec.~\ref{vacuumNJL}.
Then in Sec.~\ref{mediumQMCmodel}, the QMC model is briefly explained since this model will be used for 
estimating modified quark properties in nuclear medium.
Section~\ref{in-medium_pion} is devoted to the discussions on the obtained in-medium pion properties and
our numerical results on the change of pion properties as functions of nuclear matter density are presented
and compared with predictions of other approaches.
We then consider the quark propagator in nuclear medium and estimate the medium modifications 
of the pion elastic form factor in the combined approach of the NJL and QMC models. 
Our results on the in-medium form factor and charge radius are shown in Sec.~\ref{mediumFormFactor}, 
and we summarize in Sec.~\ref{summary}.

\section{Pion Properties in the NJL model}
\label{vacuumNJL}

In this section, we briefly review how we describe the pion in the framework of the NJL model referring 
the details to, for example, Refs.~\cite{NBC14,CBT14,HCT16}.
The NJL model is a chiral effective theory that mimics many of the key features of quantum 
chromodynamics (QCD) and, therefore, it provides a useful tool to understand non-perturbative phenomena in low energy 
QCD~\cite{BM88,KLVW90,BJM88,TTKK90,Klevansky92}. 
The two-flavor NJL Lagrangian with a four fermion contact interaction reads~\cite{CBT14}%
\footnote{In principle, the two flavor singlet pieces of the $G_\rho$ term in Eq.~\eqref{eq:lagNJL} can appear in 
the NJL interaction Lagrangian with separate coupling constants as they individually satisfy chiral symmetry. 
Our choice of identical coupling avoids flavor mixing giving the flavor content of the $\omega$ meson as 
$(u \bar{u} + d \bar{d})/\sqrt2$.}
\begin{eqnarray}
\label{eq:lagNJL}
\mathscr{L}_{\rm NJL} & = & \bar{\psi} ( i \slashed{\partial} - \hat{m} ) \psi 
+ \frac{G_\pi}{2} \left[ (\bar{\psi} \psi )^2 - (\bar{\psi} \gamma_5 \vec{\tau} \psi )^2 \right]
\nonumber \\ &&
- \frac{G_\omega}{2} (\bar{\psi} \gamma^\mu \psi )^2
- \frac{G_\rho}{2} \left[ (\bar{\psi} \gamma^\mu  \vec{\tau} \psi )^2
+ (\bar{\psi} \gamma^\mu \gamma_5 \vec{\tau} \psi )^2 \right]  ,
\nonumber \\
\end{eqnarray}
where $\vec{\tau}$ are the Pauli isospin matrices.
The quark field is an iso-doublet written as $\psi = (u, d)^T$ and $\hat{m} = \mbox{diag} (m_u, m_d)$ 
denotes the current quark mass matrix.
The four-fermion coupling constants are represented by $G_\pi$, $G_\rho$, and $G_\omega$. 
Throughout the present work, we assume $m_u = m_d = m_q$.

The general solution to the standard NJL gap equation has the form of
\begin{equation}
S^{-1}_q(p) = \slashed{p} -M_q +i \epsilon,
\end{equation}
where $q = (u, d)$ and the dressed quark mass $M_q$ is given by
\begin{eqnarray}
\label{eq:masNJL}
M_q &=& m_q - 4 \, G_\pi \braket{ \bar{q} q }
\nonumber \\
&=& m_q^{} + 12 i G_\pi \int \frac{d^4 k}{(2\pi)^4} \mbox{Tr}_D [S_q (k)].
\end{eqnarray}
Here, the quark condensate is denoted by $\braket{ \bar{q} q }$ and the trace is taken over Dirac indices only.
Introducing the proper-time regularization scheme leads to
\begin{align}
  \label{eq:massNJLProp}
  M_q &= m_q^{} + \frac{3G_\pi M_q}{\pi^2}
  \int_{1/\Lambda_{\rm UV}^{2}}^{1/\Lambda_{\rm IR}^{2}} \frac{d\tau}{\tau^2} \exp\left(-\tau 
  M_q^2\right),
\end{align}
where $\Lambda_{\rm UV}^{}$ and $\Lambda_{\rm IR}^{}$ are cutoff parameters.

The mesons considered here -- $\pi$, $\rho$, and $\omega$ -- are realized in the NJL model as 
quark-antiquark bound states whose properties are determined by the Bethe-Saltpeter equation (BSE). 
The solution to the BSE in the $\pi$ and $\beta$ ($ = \rho,\,\omega$) channels are given by a two-body 
$t$-matrix that depends on the interaction channel.
The reduced $t$-matrices in these channels read
\begin{align}
\label{eq:tmatrix}
\tau_\pi^{} (q) &= \frac{-2i\,G_\pi}{1 + 2\,G_\pi\,\Pi_\pi (q^2)}, \\
\tau_\beta^{\mu \nu} (q) &= \frac{-2i\,G_\rho}{1 + 2\,G_\rho\,\Pi_\beta (q^2)} \left(g^{\mu \nu} + 
2\,G_{\rho}\,\Pi_\beta(q^2)\, \frac{q^{\mu} 
q^{\nu}}{q^2} \right),
\end{align}
where the bubble diagrams give
\begin{align}
\label{eq:bubblegraphtot}
& \Pi_{\pi}(q^2) = 6i \int \frac{d^4k}{(2\pi)^4}\ \mbox{Tr}_D \left[\gamma_5\,S_{q}(k) \gamma_5\,S_{q}
(k+q) \right], \\
& \Pi_{\beta}^{qq} (q^2)\, P_T^{\mu\nu} = 6i \int \frac{d^4 k}{(2\pi)^4}\ \mbox{Tr}_D \left[\gamma^\mu 
S_{q} (k) \gamma^\nu S_{q}(k+q) \right],
\end{align}
with $\Pi_\rho = \Pi_\omega = \Pi_{\beta}^{qq}$ and  $P_T^{\mu\nu} = g^{\mu\nu} - q^\mu q^\nu/q^2$.

The meson masses are identified by the pole positions in the corresponding $t$-matrices, and, for example, 
the pion mass is determined by the pole condition as
\begin{align}
1 + 2\, G_\pi\, \Pi_{\pi} (q^2 = m_{\pi}^2) &= 0
\end{align}  
and analogous conditions determine $m_\rho$ and $m_\omega$. 
This procedure gives the pion mass as 
\begin{align}
  \label{eq:pionmassNJL}
m_{\pi}^2 &= \frac{m_q}{M_q} \frac{2}{G_\pi\, \mathcal{I}_{qq}(m_\pi^2)}, 
\end{align} 
where
\begin{align}
\mathcal{I}_{ab}(q^2) &= \frac{3}{\pi^2} \int_0^1 dx \int_{1/\Lambda^2_{\rm UV}}^{1/\Lambda^2_{\rm IR}} \frac{d\tau}{\tau}\,
e^{-\tau[x(x-1)\,q^2 + x\,M_b^2 + (1-x)\,M_a^2]}
\end{align} 
for quark flavor $a$ and $b$.
The residue at a pole in the $\bar{q}q$ $t$-matrix defines the effective pion-quark coupling constant $g_{\pi q q}^{}$ as
\begin{align}
\label{eq:couplinconstant}
g_{\pi q q}^{2} &= -\left[ \left.\frac{\partial\, \Pi_\pi (q^2)}{\partial q^2} \right|_{q^2 = m_\pi^2} 
\right]^{-1} .
\end{align}

The pion decay constant is determined from the meson to hadronic vacuum matrix element 
$\braket{ 0 \mid j_a^{5 \mu} (0) \mid \pi (p) }$ with $j_a^{5 \mu}$ being the weak axial-vector current operator 
for a flavor quantum number $a$. 
Solving the corresponding matrix, the pion weak decay constant in the proper-time regularization is 
obtained as~\cite{NBC14}
\begin{align}
  \label{eq:decayconNJL}
  f_\pi &= \frac{3 g_{\pi qq}^{} M_q}{4 \pi^2} \int_0^1 dx\,
  \int_{1/\Lambda_{\rm UV}^{2}}^{1/\Lambda_{\rm IR}^{2}} \frac{d\tau}{\tau} e^{-\tau [ M_q^2 - x (1-x) 
  m_\pi^2 ]}.
\end{align}
This completes the formulas for the pion decay constant and pion-quark coupling constant.
Following Ref.~\cite{CBT14}, we fix the parameter values as $M_q = 0.4$~GeV,
$\Lambda_{\rm IR} = 0.240$~GeV by the QCD scale, 
the empirical values for the pion mass ($m_\pi^{} = 140$~MeV) and the pion decay constant 
($f_\pi = 93$~MeV) as well as vector meson masses ($m_\rho = 770$~MeV and $m_\omega = 782$~MeV), 
which give
$\Lambda_{\rm UV} = 0.645$~GeV, $G_\pi = 19.0~\mbox{GeV}^{-2}$,
$G_\rho = 11.0~\mbox{GeV}^{-2}$, and $G_\omega = 10.4~\mbox{GeV}^{-2}$.
For the current quark mass, we choose $m_q^{} = 16.4$~MeV.

\section{In-medium pion properties based on the QMC model} 
\label{mediumQMCmodel}

In order to calculate in-medium $f_\pi$ and $g_{\pi qq}$ in the framework of the NJL model, we need the 
medium modifications of the quark properties. 
For this purpose, we adopt the QMC model for the in-medium modified quark properties in nuclear matter.
The QMC model~\cite{Guichon88} has been successfully applied to many phenomena of nuclear and hadron 
systems including finite nuclei~\cite{GST18,SGRT16,GSRT95,STT96b,STT96}, hypernuclei~\cite{TSHT98,GTT07}, 
superheavy nuclei~\cite{SGT17}, and neutron stars~\cite{WCTTS13}.
It has also been widely used to study the in-medium nucleon and hadron 
properties~\cite{STT05,KTT17}.  
Recently, studies have been extended to the calculation of the medium modifications of the nucleon weak and 
electromagnetic form factors on the neutrino mean free path in dense matter~\cite{HOT18}.
In this section, we briefly review the main features of this model.

In the QMC model, medium effects arise from the self-consistent exchange of the scalar ($\sigma$) and 
vector ($\omega$ and $\rho$) meson fields directly coupled to the confined quarks rather than to the pointlike nucleon. 
In the present calculation, we consider symmetric nuclear matter in the Hartree mean field 
approximation. 
This is because the differences between the results of the Hartree and Hartree-Fock calculations are 
found to be relatively small. 
In particular, the energy densities per nucleon for symmetric nuclear matter are nearly identical at the 
cost of complications introduced
in the Hartree-Fock treatment. 
Therefore, we use the Hartree approximation in this exploratory study. 
More details on the Fock terms in the QMC model can be found, for example, in Ref.~\cite{KTT98}.

The effective Lagrangian for symmetric nuclear matter in the QMC model is given 
by~\cite{GSRT95,STT96b,STT96,TSHT98,GTT07,STT05}
\begin{align}
  \label{eqintro1}
\mathscr{L}_{\rm QMC} & =  \bar{\psi} \left[ i \gamma_\mu \partial^\mu - M_{N}^{*}(\sigma) 
- g_\omega^{} \gamma_\mu \omega^{\mu} \right] \psi + \mathscr{L}_\textrm{meson},
\end{align}
where $\psi$, $\sigma$, and $\omega$ are the nucleon, isoscalar-scalar meson $\sigma$, and 
isoscalar-vector meson $\omega$ fields, respectively.
The effective mass of the nucleon $M_N^{*} (\sigma)$ is defined as
\begin{align}
  \label{eqintro2}
M_{N}^{*} \left(\sigma \right) &\equiv M_{N} 
- g_\sigma^{} \left( \sigma \right) \sigma,
\end{align}
where $g_{\sigma} \left( \sigma \right)$ and $g_{\omega}$ are the $\sigma$-dependent nucleon$-\sigma$ 
coupling and
the nucleon$-\omega$ coupling constants, respectively. 
In symmetric nuclear matter, the isospin-dependent $\rho$-meson field vanishes in the Hartree 
approximation, so we do not explicitly include the $\rho$ meson field in the present work.
The mesonic part of the Lagrangian in Eq.~(\ref{eqintro1}) reads
\begin{eqnarray}
\label{eqintro3}
\mathscr{L}_\textrm{meson} &=& \frac{1}{2} \left( \partial_\mu \sigma \partial^\mu \sigma - m_\sigma^2 
\sigma^2 \right) 
- \frac{1}{2} \partial_\mu \omega_\nu \left( \partial^\mu \omega^\nu - \partial^\nu \omega^\mu \right)
\nonumber \\ && \mbox{}
+ \frac{1}{2} m_\omega^2 \omega^\mu \omega_\mu ,  
\end{eqnarray}

The nucleon Fermi momentum $k_F^{}$, nucleon (baryon) density $\rho_B^{}$, and the scalar density $\rho_{s}^{}$ 
of nuclear matter are defined as
\begin{align}
\label{eqintro4}
\rho_{B}^{} &= \frac{\gamma}{(2\pi)^3} \int d {\bf k}\, \theta ( k_F^{}  - | {\bf k} | ) = 
\frac{\gamma k_F^3}{3 \pi^2}, 
\nonumber \\
\rho_{s}^{} &= \frac{\gamma}{(2\pi)^3} \int d {\bf k} \, \theta ( k_F^{} - | {\bf k} | )
\frac{M_N^{*} (\sigma)}{\sqrt{M_N^{*2} (\sigma ) + {\bf k}^2}},
\end{align}
where $\gamma = 4$ for a symmetric nuclear matter and $\gamma = 2$ for asymmetric nuclear matter. 
The Fermi momenta of the proton and neutron $k_F^{p,n}$ are respectively determined by $\rho_p^{}$ and 
$\rho_n^{}$ with $\rho_B^{} = \rho_p^{} +\rho_n^{}$.

In the QMC model, nuclear matter is treated as a collection of the nucleons that are assumed to be 
non-overlapping MIT bags~\cite{CJJTW74}. 
The Dirac equations for the light $q$ ($u$ and $d$) quarks inside the bag are given by
\begin{align}
\label{eqintro5}
\left[ i \gamma \cdot \partial - \left( m_q^{} - V_{\sigma}^{q} \right)  \mp \gamma^{0} V_{\omega}^{q} 
\right] \left( \begin{array}{c} \psi_u(x)  \\ \psi_{\bar{u}}(x) \\ \end{array} 
\right) &= 0 , 
\nonumber \\
\left[ i \gamma \cdot \partial - \left( m_q^{} - V_{\sigma}^{q} \right) \mp \gamma^{0}  V_{\omega}^{q} 
 \right] \left( \begin{array}{c} \psi_d(x)  \\ \psi_{\bar{d}}(x) \\ \end{array} 
\right) &= 0,
\end{align}
which define the effective current quark mass $m_q^{*}$ as
\begin{align}
\label{eqintro5a}
m_q^{*} & \equiv m_q^{} - V_{\sigma}^{q} .
\end{align}
As in the previous section, we assume $m_u = m_d = m_q$.
The scalar and vector mean fields in symmetric nuclear 
matter are defined as
\begin{align}
V_{\sigma}^{q}  \equiv g_{\sigma}^{q} \braket{ \sigma }, \qquad
V_{\omega}^{q} \equiv g_{\omega}^{q} \, \delta^{\mu 0} \braket{ \omega^{\mu}}.
\label{vqpot}
\end{align}

The bag radius of the nucleon in nuclear medium ($R_N^{*}$) is determined from the mass stability condition 
against the variation of the bag radius.
The eigenenergies of quarks in units of $1/ R_N^{*}$ are given by
\begin{align}
\label{eq:pionmed9}
\left( \begin{array}{c}
 \epsilon_u \\ \epsilon_{\bar{u}}
\end{array} \right)
& = \Omega_q^* \pm R_N^* \left( V^q_\omega + \frac{1}{2} V^q_\rho \right), \nonumber \\
\left( \begin{array}{c} \epsilon_d \\  \epsilon_{\bar{d}}
\end{array} \right)
&= \Omega_q^* \pm R_N^* \left( V^q_\omega - \frac{1}{2} V^q_\rho \right),
\end{align}
and the in-medium effective mass of the nucleon is obtained as
\begin{align}
\label{eq:pionmed10}
M_N^{*} &= \sum_{j = q, \bar{q}} \frac{n_j \Omega_j^{*} -z_N^{}}{R^{*}_N} 
+ \frac{4}{3} \pi R_N^{* 3} B,
\end{align}
where $\Omega^{*}_q = \Omega^{*}_{\bar{q}} = \left[ x_q^2 + \left(R_N^{*} m_q^{*} \right)^2 \right]^{1/2}$ 
with $x_q$ being the lowest mode bag eigenvalue that will be specified later.
In this expression, $z_N^{}$ is related with the sum of the center-of-mass and gluon fluctuation corrections~\cite{GSRT95}.
Its value is determined by the nucleon mass in free space and is assumed to be independent of density~\cite{GSRT95}.
The bag pressure $B$ is also fixed by the inputs for the nucleon in vacuum, $R_N = 0.8$~fm
and the free nucleon mass $M_N=939$ MeV.
The parameter values are determined for a given value of the current quark mass $m_q^{}$ and they are summarized in Table~\ref{tab:model1}.
Then the in-medium bag radius $R_N^{*}$ is determined by the condition
\begin{align}
\label{eq:pionmed10a}
\frac{\partial M_N^{*}}{\partial R_N^*} = 0.
\end{align}
The nucleon bag radius $R^*_N$ at normal nuclear matter density is found to decrease by a few per cent~\cite{STT05}. 
It should, however, be mentioned that this is not a physical quantity and the corresponding quantity must be 
calculated using the relevant wave function.
In fact, in Ref.~\cite{STT96b}, it is shown that the nucleon radius at normal nuclear matter density increases by a few percent 
when calculated with the quark wave function.

\begin{table}[t]
\caption{Parameters of the QMC model and the obtained nucleon properties at saturation density 
$\rho_0^{} = 0.15~\mbox{fm}^{-3}$ for two quark mass values in free space, $m_q^{} = 5.0$ and $16.4$~MeV. 
The current quark mass $m_q$, the effective nucleon mass $M_N^{*}$, and the nuclear incompressibility $K$ are given 
in units of MeV. 
The parameters are fitted to the free space nucleon mass $M_N=939$~MeV with $R_N = 0.8~\mbox{fm}$, 
and the nuclear matter saturation properties, i.e., the binding energy of 15.7~MeV at the saturation density $\rho_0^{}$. }
\label{tab:model1}
\addtolength{\tabcolsep}{3.4pt}
\begin{tabular}{ccccccc} 
\hline \hline
$m_q$ & $g_{\sigma}^2 / 4 \pi$ & $g_{\omega}^2 / 4 \pi$ & $B^{1/4}$ & $z_N^{}$ & $M_N^{*}$ & $K$ \\[0.2em] 
\hline
$5.0 $   & $5.393$ & $5.304$ & $170.0$ & $3.295$ & $754.6$ & $279.3$ \\ 
$16.4$ & $5.438$ & $5.412$ & $169.2$ & $3.334$ & $752.0$ & $281.5$
\\ \hline \hline
\end{tabular}
\end{table}

All the parameter values and some results of the QMC model are displayed in Table~\ref{tab:model1} 
for two different quark mass values, $m_q^{} = 5.0$ and $16.4$~MeV. 
In the present calculation, to be consistent with the NJL model described in the previous section,
we use $m_q^{}=16.4$~MeV. 
The ground state wave function of the quark inside a bag satisfies the boundary condition at the bag surface,
$j_0 (x_q) =  \beta_q j_1 (x_q)$, where $j_0$ and $j_1$ are the spherical Bessel functions and
\begin{align}
  \label{eq:beta}
  \beta_q &= \sqrt{\frac{\Omega_q^{*} -m_q^{*} R_N^{*}}{\Omega_q^{*} + m_q^{*} R_N^{*}}} .
\end{align}

The scalar $\sigma$ and vector $\omega$ meson fields at the nucleon level are related as
\begin{align}
\label{eq:pionmed11}
\omega &= \frac{g_\omega \, \rho_B }{m_\omega^2},\\
\label{eq:sigma}
\sigma &= \frac{4 g_\sigma \, C_N (\sigma)}{(2\pi)^3m_{\sigma}^2}
\int d{\bf k} \, \theta (k_F^{} - | {\bf k} | ) 
\frac{M_N^{*} (\sigma)}{\sqrt{M_N^{*2} (\sigma) + {\bf k}^2}},
\end{align}
where $C_N (\sigma)$ is defined as
\begin{align}
  C_N (\sigma) &= \frac{-1}{g_\sigma (\sigma =0)} 
\left[ \frac{\partial M_N^{*} (\sigma )}{\partial \sigma } \right].
\end{align}
For the point-like structureless nucleon, we have $C_N (\sigma) = 1$ as in Quantum Hadrodynamics (QHD)~\cite{SW86,SW97}.
Thus the $\sigma$ dependence of the coupling $g_\sigma (\sigma)$ or $C_N(\sigma)$ reflects the modifications of the quark dynamics in nuclear medium 
and is the origin of the novel saturation properties achieved in the QMC model.
By solving the self-consistent equation for the scalar $\sigma$ mean field in Eq.~(\ref{eq:sigma}), 
the total energy per nucleon is obtained as
\begin{align}
  \label{eq:pionmed12}
  E^{\rm tot}/A &= \frac{4}{(2\pi)^3 \rho_B^{}} \int d {\bf k} \, \theta (k_F^{} - | {\bf k} |) \sqrt{M_N^{*2} (\sigma) 
+ {\bf k}^2} \nonumber \\
&+ \frac{m_\sigma^2\, \sigma^2}{2\rho_B^{}} 
+ \frac{g_\omega^2\, \rho_B^{}}{2 m_\omega^2}.
\end{align}
The coupling constants $g^q_\sigma$ and $g^q_\omega$ in Eq.~(\ref{vqpot}) are determined by fitting the 
binding energy, 15.7~MeV, of symmetric nuclear matter at the saturation density 
and they are respectively related with $g_\sigma$ and $g_\omega$ by $g_\sigma^{} = 3 S_N(\sigma=0) g^q_\sigma$ and 
$g_\omega^{} = 3 g^q_\omega$, where $S_N(\sigma)$ is defined through~\cite{Guichon88,KTT17}
\begin{eqnarray}
\dfrac{\partial M_{N}^*(\sigma)}{\partial \sigma}
&=& - 3 g_{\sigma}^q \int_{\rm bag} d\,{\bf y} \ {\overline \psi}_q({\bf y})\, \psi_q({\bf y})
\nonumber \\
&\equiv& - 3 g_{\sigma}^q S_{N}(\sigma) = - \dfrac{\partial}{\partial \sigma}
\left[ g^{N}_\sigma(\sigma) \sigma \right]
\label{Ssigma}
\end{eqnarray}
with $\psi_q$ being the lowest mode bag wave function in medium.
These relations determine the in-medium quark dynamics, which is needed to compute the medium modifications 
of \textit{hadron} properties in nuclear medium.
That is, we assume that the in-medium light-quark properties obtained for the nucleon are not much different from those
of light quarks in other hadrons. 
In the present work, therefore, we will explore the in-medium pion properties with the modified quark properties determined by the 
in-medium nucleon properties.

\begin{figure}[t]
\centering\includegraphics[width=0.95\columnwidth]{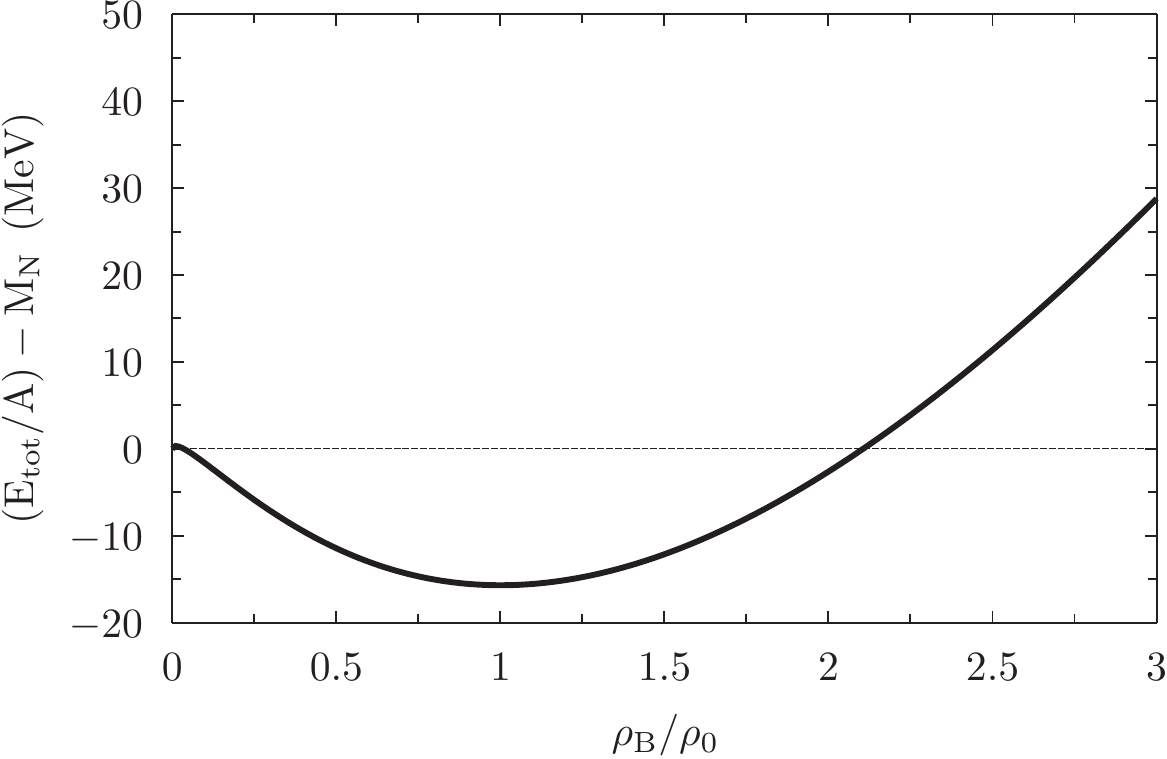}
\caption{\label{fig1} 
Energy per nucleon, $E^{\rm tot} /A - M_N $, for symmetric nuclear matter in the QMC model for
the current quark mass $m_q = 16.4$~MeV.
}
\end{figure}

\begin{figure}[t]
\centering\includegraphics[width=0.95\columnwidth]{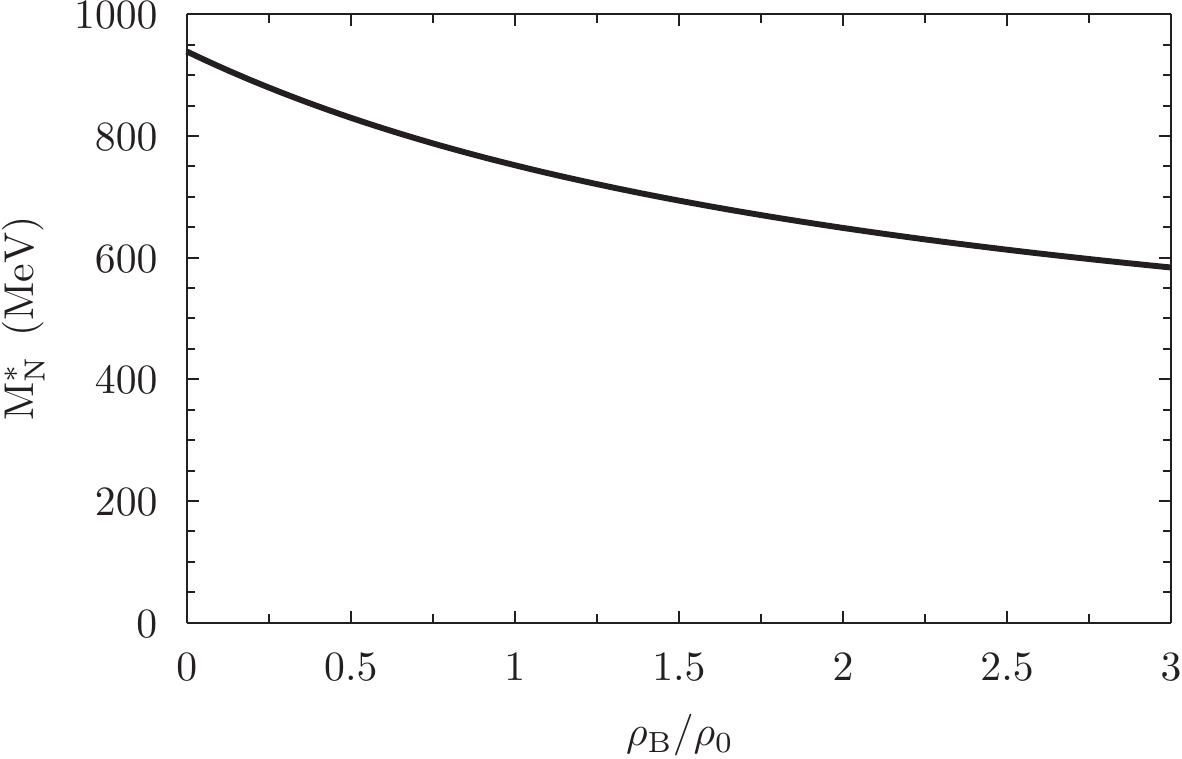}
\caption{\label{fig2} 
Effective nucleon mass $M^{*}_N$ in symmetric nuclear matter calculated with $m_q = 16.4 $~MeV. 
}
\end{figure}

\begin{figure}[t]
\centering\includegraphics[width=0.95\columnwidth]{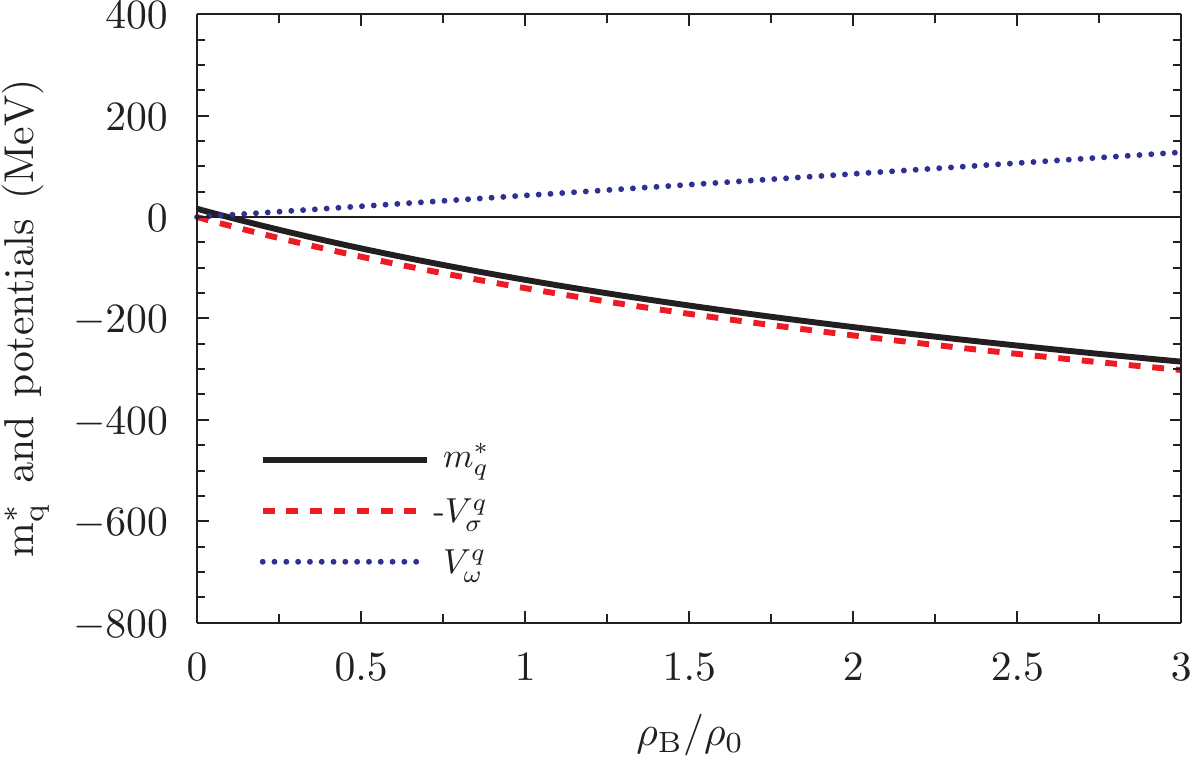}
\caption{\label{fig3} 
The effective quark mass, $m^{*}_q$, and the quark potentials, $-V^q_{\sigma}$ and $V^q_{\omega}$, for $m_q = 16.4 $~MeV. 
}
\end{figure}

Figure~\ref{fig1} shows the energy per nucleon for symmetric nuclear matter, which corresponds to the results 
listed in Table~\ref{tab:model1} for $m_q = 16.4$~MeV.  
The obtained effective nucleon mass $M^*_N$ is illustrated in Fig.~\ref{fig2} as a function of $\rho_B^{} / \rho_0^{}$.
The corresponding effective quark mass $m^*_q$ as well as the scalar potential $-V^q_\sigma$ and vector potential 
$V^q_\omega$ are shown in Fig.~\ref{fig3}.

\section{In-medium pion properties}
\label{in-medium_pion}

\begin{figure}[t]
\centering\includegraphics[width=0.95\columnwidth]{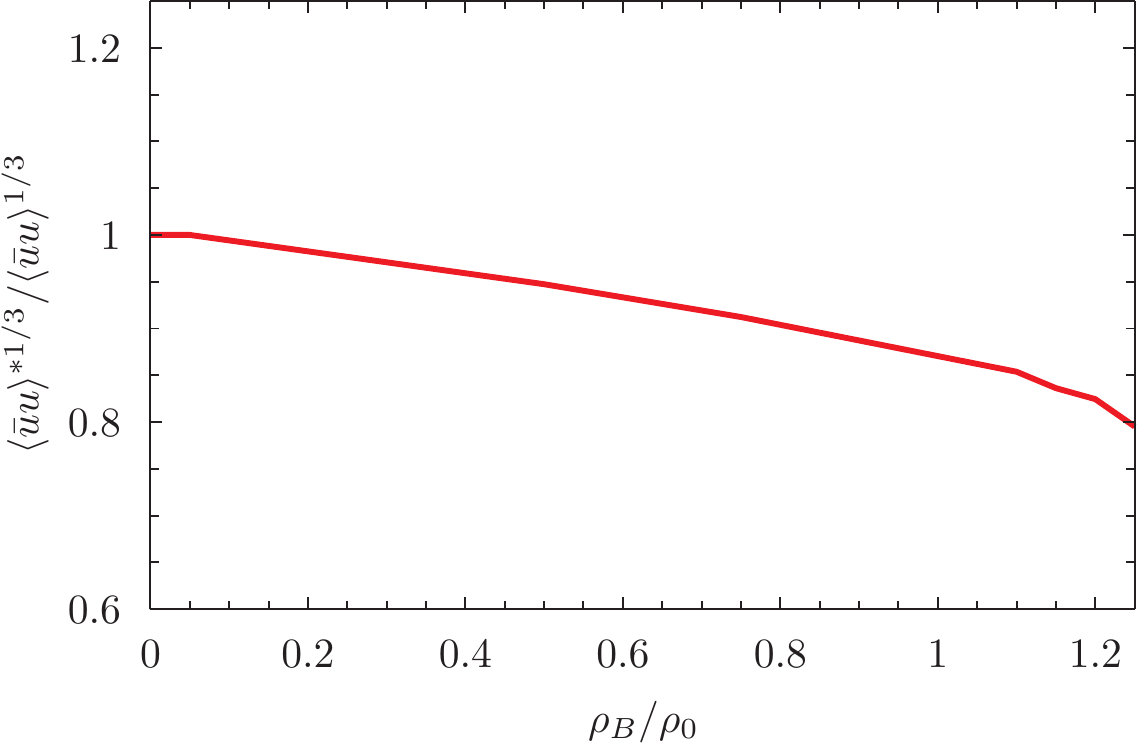}
\caption{\label{fig4} 
The ratio of the in-medium to vacuum quark condensate as a function of $\rho_B^{} / \rho_0^{}$.}
\end{figure}

\begin{figure}[t]
\centering\includegraphics[width=0.95\columnwidth]{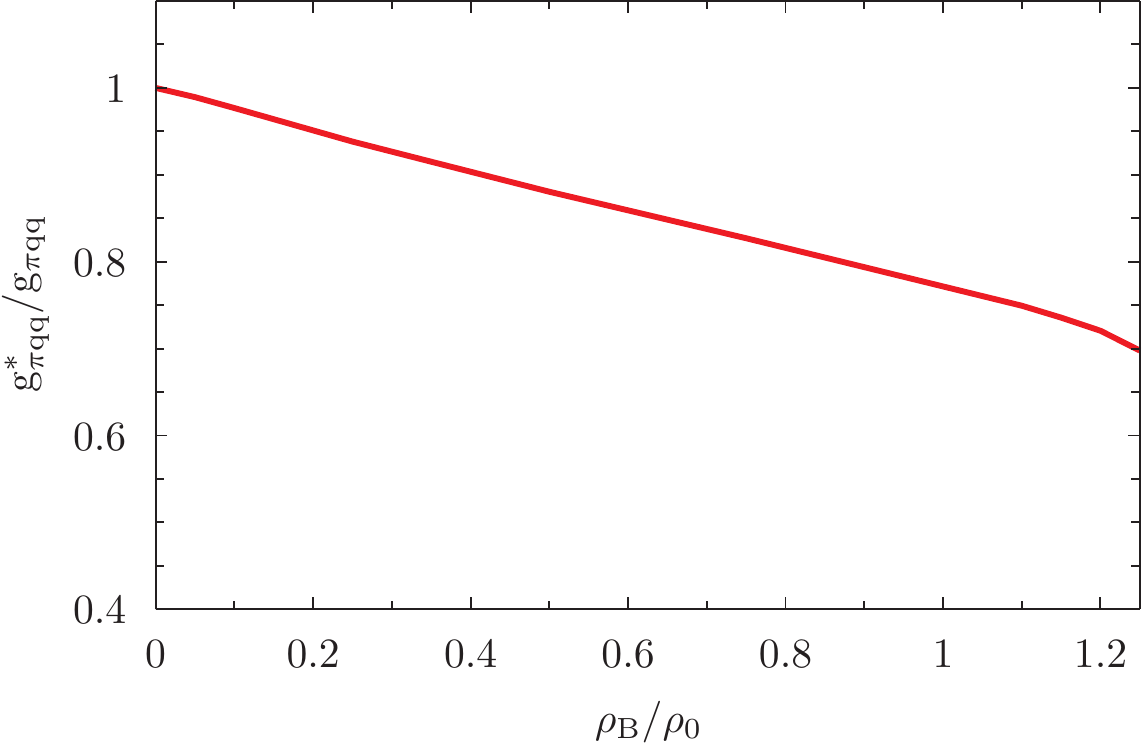}
\caption{\label{fig5} 
The ratio of the in-medium to vacuum pion-quark-quark 
coupling constant as a function of $\rho_B^{} / \rho_0^{}$.} 
\end{figure}

\begin{figure}[t]
\centering\includegraphics[width=0.95\columnwidth]{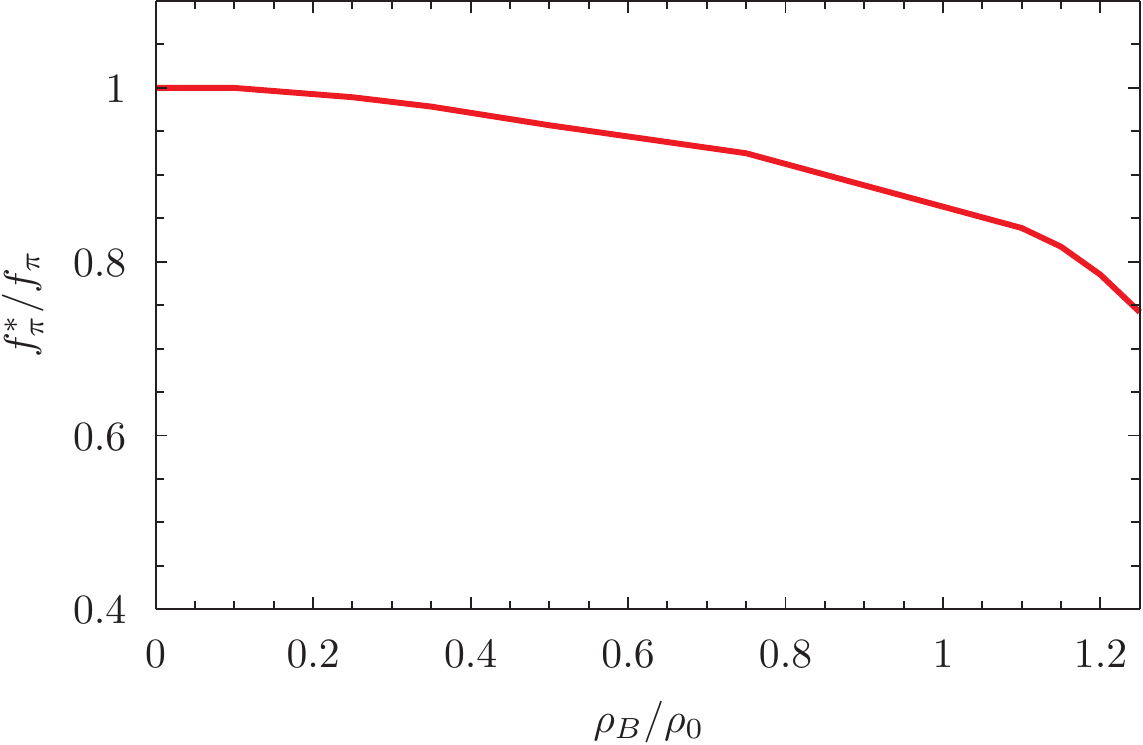}
\caption{\label{fig6} 
The ratio of the in-medium to vacuum pion decay constant as a function of $\rho_B^{} / \rho_0^{}$.} 
\end{figure}

\begin{figure}[t]
\centering\includegraphics[width=0.95\columnwidth]{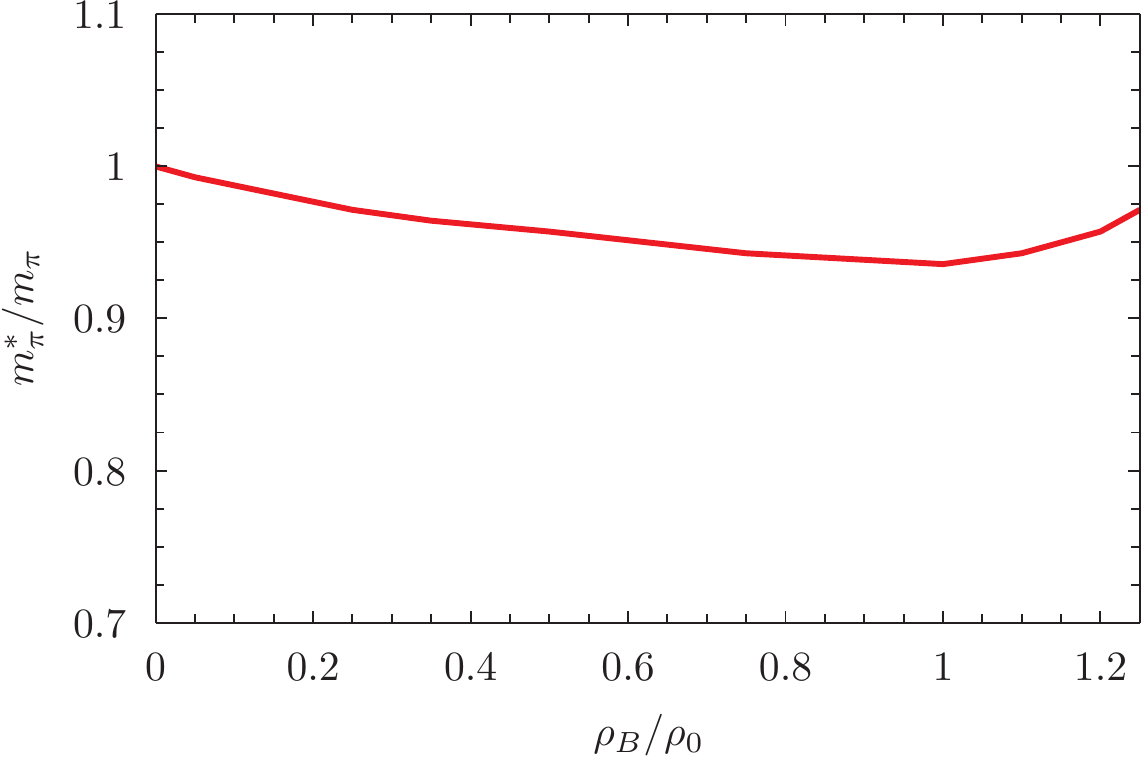}
\caption{\label{fig7} 
The ratio of the in-medium to vacuum pion mass as a function of $\rho_B^{} / \rho_0^{}$.} 
\end{figure}

Being equipped with the NJL-model formalism and QMC model for the medium-modified quark properties, 
we compute the pion properties in nuclear medium in this section.
Using the in-medium properties obtained in the QMC model with $m_q = 16.4$~MeV, we calculate the effective 
quark mass $M^*_u$, in-medium pion mass, in-medium pion decay constant, in-medium quark condensate, and in-medium 
$\pi q q$ coupling constant.
The results are listed in Table~\ref{tab:model2} for a given set of nuclear matter densities.
Some of these results are illustrated in Figs.~\ref{fig4}--\ref{fig7} as functions of $\rho_B^{} / \rho_0^{}$.

\begin{table}[t]
\caption{Properties of in-medium pion calculated in the NJL model with the in-medium inputs from the QMC model for
$m_q = 16.4$ MeV. The units for $M_u^*$, $m_\pi^*$, $f_\pi^*$, and $- \braket{ \bar{u} u }^{* 1/3}$ are GeV.
}
\label{tab:model2}
\addtolength{\tabcolsep}{4.6pt}
\begin{tabular}{cccccc} 
\hline \hline
$\rho_B^{} / \rho_0^{}$ & $M_u^{*} $ & $m_\pi^{*} $ & $f_\pi^{*}$ 
& $g_{\pi q q}^{*}$ & $- \braket{ \bar{u} u }^{* 1/3}$  \\[0.2em] 
\hline
$0.00$ & $0.400$ & $0.140$ & $0.093$ & $4.225$ & $0.171$ \\ 
$0.25$ & $0.370$ & $0.136$ & $0.092$ & $3.964$ & $0.167$ \\
$0.50$ & $0.339$ & $0.134$ & $0.089$ & $3.720$ & $0.162$ \\
$0.75$ & $0.307$ & $0.132$ & $0.086$ & $3.494$ & $0.156$ \\
$1.00$ & $0.270$ & $0.131$ & $0.081$ & $3.265$ & $0.149$ \\
$1.25$ & $0.207$ & $0.136$ & $0.069$ & $2.948$ & $0.136$
\\ \hline \hline
\end{tabular}
\end{table}

Shown in Fig.~\ref{fig4} are the results for the ratio of the in-medium to vacuum quark condensates.
We found that this ratio decreases with increasing nuclear matter density and the ratio at 
the nuclear matter saturation density is estimated to be about 0.87.
This is somehow higher than the value obtained in Ref.~\cite{JHK08}, 
which gives $0.63 \sim 0.57$ through the
relation $ \braket{ \bar{q} q }^{*} / \braket{ \bar{q} q } \sim 1 - (0.37 \sim 0.43)\,  
\rho_B^{} / \rho_0^{}$ in low density region.
Results for the ratio of the in-medium to vacuum pion-quark coupling constant $g^*_{\pi qq}/g_{\pi qq} $ 
are shown in Fig.~\ref{fig5}. 
We again observe that $g_{\pi q q}^{*} / g_{\pi q q}$ decreases with increasing density,
which is consistent with the results of Refs.~\cite{BM88d,RTT12}.
At normal nuclear density, we obtain $g^*_{\pi qq}/g_{\pi qq} = 0.77$, which is smaller than the value ($\approx 0.9$) 
obtained in Ref.~\cite{RTT12}.

In the NJL model we calculate the in-medium pion decay constant $f_\pi^*$ with Eq.~(\ref{eq:decayconNJL}) 
by replacing quark mass and coupling by the corresponding in-medium quantities.%
\footnote{In a nuclear medium, the pion decay constant can be separated into the temporal $f_\pi^{(t)}$ and spatial $f_\pi^{(s)}$ components, respectively,
where the $f_\pi^{(t)}$ is directly connected to a physical quantity~\cite{KW97b}.}
Our results for the ratio of the in-medium to vacuum pion decay constants are presented in Fig.~\ref{fig6}. 
Again the ratio is found to decrease as density increases, and, at normal nuclear matter density, 
we obtain $f_\pi^{*} / f_\pi = 0.87$, which is in good agreement with the results of Refs.~\cite{KW97b,KY03}, 
$f_\pi^{*} / f_\pi = 0.80$, and is larger than the values obtained in Refs.~\cite{MOW02,TW95} by about 10-20\%.

In vacuum or at zero baryon density, the NJL model satisfies the chiral limit as can be seen from Eq.~\eqref{eq:pionmassNJL}.
Namely, in the chiral limit, $m_q = 0$, where $m_q$ is the current quark mass, the pion mass vanishes, although 
the constituent quark mass $M_q \ne 0$ since the chiral condensate does not vanish as shown in Eqs.~\eqref{eq:masNJL} 
and \eqref{eq:massNJLProp}.
By combining the result of the gap equation of Eq.~\eqref{eq:massNJLProp} with the pion mass and the pion decay constant 
given in Eqs.~\eqref{eq:pionmassNJL} and \eqref{eq:decayconNJL}, respectively, it is straightforward to obtain
the Gell-Mann--Oakes--Renner relation,
\begin{eqnarray}
  \label{eq:condensate}
  f_\pi^2 m_\pi^2 &\equiv& - \frac{1}{2} (m_u + m_d) \langle \bar{u} u + \bar{d} d \rangle,
\end{eqnarray}
where $\langle \bar{u} u \rangle$ and $\langle \bar{d} d \rangle$ are the light $u$- and $d$-quark (chiral) condensates, 
respectively, which are directly related to the constituent quark mass in the chiral limit.
This explicitly shows that the present proper-time regularization scheme in the NJL model is consistent with the chiral symmetry in QCD.

At finite baryon density, the in-medium quark condensate $\langle \bar{q} q \rangle^*$, which is the order parameter of chiral symmetry 
in medium, also can be calculated similarly to the vacuum case with the corresponding in-medium inputs using
Eqs.~\eqref{eq:masNJL} and \eqref{eq:massNJLProp}. 
The calculated ratio for the in-medium to vacuum light quark condensate as a function of $\rho_B^{} / \rho_0^{}$ is shown in Fig.~\ref{fig4}. 
The result shows the decrease of the in-medium light quark condensate as nuclear matter density increases. 
This indicates that chiral symmetry is partially restored at finite baryon density or by increasing nuclear matter density.

Finally, we check the medium modifications on the pion mass and the results are presented in Fig.~\ref{fig7}.
We confirm that the pion mass is almost unchanged up to $1.25\, \rho_0^{}$,
which is consistent with the results obtained in Ref.~\cite{BM88d} in the low density region. (See also Ref.~\cite{KW97b}.)
The differences between the in-medium pion mass and the free pion mass is within 6\% up to nuclear density of $1.25\,\rho_0^{}$.
This justifies our assumption that $m_\pi^* \approx m_\pi$ up to about normal nuclear density.

With the in-medium modifications on the pion properties obtained in this section,
we also consider the quenching of the nucleon weak axial-vector 
coupling constant.
The original Goldberger-Treiman relation is at the nucleon level and is given by $g_A^{} = g_{\pi N N}^{} {f_\pi}/{M_N}$, 
where $g_A$, $g_{\pi N N}^{}$, and $M_N$ are the isovector nucleon weak axial-vector coupling constant, 
pion-nucleon coupling, and the nucleon mass, respectively.
At the quark level, $M_N$ is replaced by $M_q$ (dynamical quark mass) and $g_{\pi N N}^{}$ by $g_{\pi q q}^{}$. 
Then, the ratio of the in-medium to vacuum isovector nucleon axial-vector coupling constant can be written as~\cite{RTT12}
\begin{align}
 \label{eq:gmt}
 \left(\frac{g_{A}^{*}}{ g_{A}} \right) 
&= \left( \frac{g_{\pi q q}^{*}}{g_{\pi q q}} \right) 
\left( \frac{f_\pi^{*}}{f_\pi} \right) \left( \frac{M_q}{M_q^{*}} \right).  
\end{align}
By inserting our results for the corresponding quantities, we estimate $g_A^{*} / g_A \approx 0.99$, 
which is consistent with the quenching of $g_A^{*}$ but \textit{less amount of quenching\/} compared with the result of
Ref.~\cite{LTT01} that gives $g_A^{*} / g_A \approx 0.9$.

\section{In-Medium Electromagnetic Form Factors} \label{mediumFormFactor}

In this section we calculate the electromagnetic form factor of the positively charged in-medium pion.
The in-medium pion electromagnetic form factors are determined by adopting 
the method of Ref.~\cite{HCT16} with a dressed quark-photon vertex.

\subsection{Formalism}

The in-medium electromagnetic form factor is obtained by modifying the quark propagator as  
\begin{equation}
  \label{eq:prognjl}
S_q^{*}(k^{*}) =  \frac{1}{ \slashed{k}^{*} - M_q^{*} + i \epsilon },
\end{equation} 
where $M_q^{*}$ is the effective constituent quark mass and the in-medium quark momentum is given   
by $k^{*}_\mu = k_\mu + V_\mu$ with vector field $V_\mu=(V_0,{\bf 0})$ by neglecting the modification 
of the space component of the quark momentum that is known to be small~\cite{KTT98,MKS11}.
In the form factor calculation, the vector field, which enters the propagator in Eq.~\eqref{eq:prognjl},
can be eliminated by the shift of the variable in the integration~\cite{DTERF14,CMPR09}.

The electromagnetic current of the in-medium pion is given by
\begin{align}
\label{eq:formfactor1}
J^{\mu} (p',p) &= \left(p'^\mu + p^\mu \right) F_\pi^{*} (Q^2),
\end{align}
where $p$ and $p'$ are the initial and final four momenta of the pion with $q^2 =(p'-p)^2 \equiv -Q^2$
and $F_\pi^{*} (Q^2)$ is the in-medium pion form factor.

\begin{figure}[t]
\centering\includegraphics[width=\columnwidth]{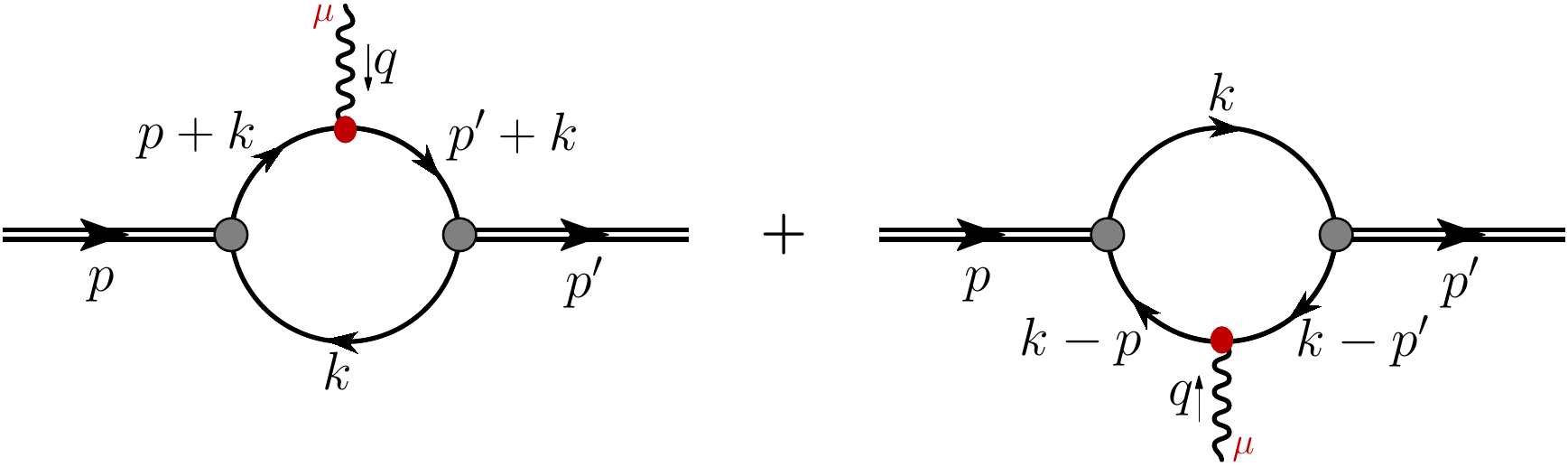}
\caption{Electromagnetic interactions of the pion.}
\label{fig:emvertex1}
\end{figure}

As in the case of the pion in vacuum~\cite{HCT16}, the in-medium pion form factor in the NJL model 
is written as the sum of the two diagrams depicted in Fig.~\ref{fig:emvertex1}, which respectively give 
\begin{widetext}
\begin{align}
\label{eq:j1}
j^{\mu}_{1,\pi}\left(p’,p\right) &= i \left( g_{\pi qq}^{*} \right)^2 \int \frac{d^4k}{(2\pi)^4}
\mathrm{Tr}\left[\gamma_5\,\tau_\alpha^\dagger\,S^{*} (p'+k)\,\hat{Q}\,\gamma^\mu\,S^{*} (p+k)\,
\gamma_5\,\tau_\alpha\,S^{*}(k)\right], \\
\label{eq:j2}
j^{\mu}_{2,\pi}\left(p’,p\right) &= i \left( g_{\pi qq}^{*} \right)^2 \int \frac{d^4k}{(2\pi)^4} 
\mathrm{Tr}\left[\gamma_5\,\tau_\alpha\,S^{*} (k-p)\,\hat{Q}\,\gamma^\mu\,S^{*} (k-p')\,
\gamma_5\,\tau_\alpha^\dagger\,S^{*}(k)\right],
\end{align}
\end{widetext}
where $\hat{Q} = \frac{1}{6} + \frac{\tau_3}{2}$ and the trace is taken over the Dirac, color, and flavor indices. 
The index $\alpha$ labels the state and $\tau_\alpha = \frac{1}{\sqrt{2}} (\tau_1 \pm i \tau_2 )$ are 
the corresponding flavor matrices. 
In flavor space, the in-medium quark propagator reads $S_q^{*}(p) = \text{diag}[S^{*}_u(p),\,S^{*}_d(p)]$.

The in-medium form factor of the pion is then given by
\begin{align}
\label{eq:bareKplus}
F_{\pi^{+}}^{\text{* (bare)}}(Q^2) &= \left( e_u - e_d \right) \, f^{* \ell \ell}_\pi (Q^2) ,
\end{align}
where $\ell = u, d$.
The superscript ``(bare)'' in Eq.~(\ref{eq:bareKplus}) means that the quark-photon vertex is elementary, that is,  
$\Lambda_{\gamma q}^{\mu\text{(bare)}} = \hat{Q}\,\gamma^\mu$. 
The first superscript $a$ in $f_\pi^{* ab}(Q^2)$ indicates the struck quark 
and the second superscript $b$ means the spectator quark.
The explicit expression of $f_\pi^{* ab}(Q^2)$ reads
\begin{widetext}
\begin{eqnarray}
f^{* ab}_\pi (Q^2) &=& \frac{3 \left( g_{\pi qq}^{*} \right)^2 }{4\,\pi^2} 
\int_0^1 dx \int_{1/\Lambda^2_{\rm UV}}^{1/\Lambda^2_{\rm IR}} 
\frac{d\tau}{\tau}\ \exp \{-\tau\left[M_a^{* 2} + x(1-x)\,Q^2\right] \} 
\nonumber \\  && \mbox{}
+ \frac{3  \left( g_{\pi qq}^{*} \right)^2}{4\,\pi^2}  \int_0^1\! dx\! \int_0^{1-x}\! dz\! 
\int_{1/\Lambda^2_{\rm UV}}^{1/\Lambda^2_{\rm IR}}  \! d\tau \Bigl[(x+z)\,m_\pi^{* 2} 
+ (M^{*}_a - M^{*}_b)^2(x+z) + 2 M^{*}_b\left(M^{*}_a - M^{*}_b\right)\Bigr] 
\nonumber \\ &&  \mbox{} \qquad \qquad
\times \exp \{-\tau\left[(x+z)(x+z-1)\,m_\pi^{* 2} + (x+z)\,M_a^{* 2} 
+ (1-x-z)\,M_b^{* 2} + x z Q^2\right] \}.
\end{eqnarray}
\end{widetext}

\begin{figure}[t]
\centering\includegraphics[width=\columnwidth]{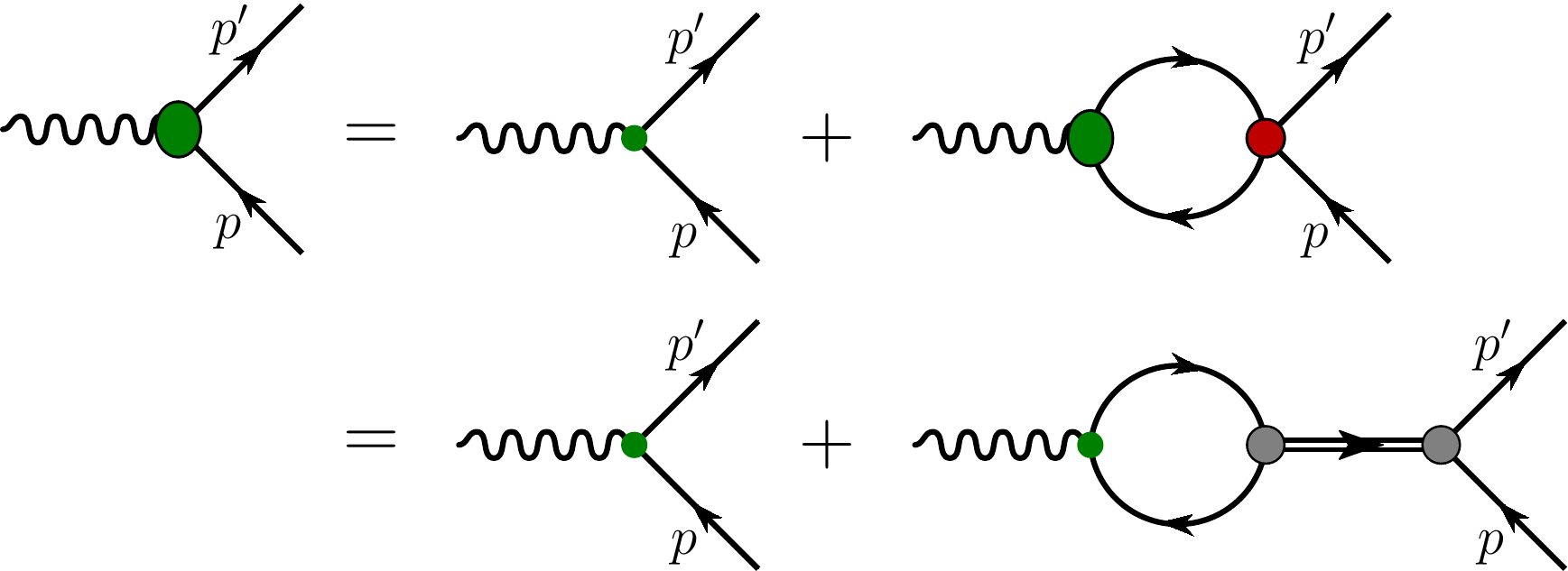}
\caption{Diagrams of the inhomogeneous Bethe-Salpeter equation that gives the dressed quark-photon vertex.  
The large shaded oval represents the solution to the inhomogeneous BSE, while the small dot is the inhomogeneous 
driving term $\hat{Q}\,\gamma^\mu$, and the double-dots represent the $q\bar{q}$ interaction kernel.}
\label{fig:vectormesons}
\end{figure}

In general, the quark-photon vertex is not simply given as $\hat{Q}\,\gamma^\mu$, but is dressed, of which effect is given by 
the inhomogeneous Bethe-Salpeter equation as illustrated in Fig.~\ref{fig:vectormesons}~\cite{HCT16}. 
With the NJL interaction kernel, the general form for the dressed quark-photon vertex for a quark of flavor $f$ reads
\begin{eqnarray}
\label{eq:quarkvertex}
\Lambda^\mu_{\gamma\,Q}(p’,p) &=& e_f^{} \gamma^\mu 
+ \left(\gamma^\mu - \frac{q^\mu\slashed{q}}{q^2}\right)F^{*}_{Q}(Q^2)
\nonumber \\ & \to &
\gamma^\mu\,F^{*}_{1Q}(Q^2),
\end{eqnarray}
where the final form follows from that the $q^\mu\slashed{q}/q^2$ term does not contribute because of the transversality of the photon. 
One can verify that, after the substitution in Eq.~\eqref{eq:quarkvertex}, it clearly satisfies the Ward-Takahashi identity, 
$q_\mu\,\Lambda^\mu_{\gamma\,Q}(p’,p) = e_f^{} \left[S_q^{* -1}(p') - S_q^{* -1}(p)\right]$. 
For the dressed $u$ and $d$ quarks we find
\begin{align}
\label{eq:f1U}
F^{*}_{1U/D}(Q^2) &= e_{u/d}\ \frac{1}{1 + 2\,G_{\rho(\omega)} \,\Pi_{\beta}^{* \ell\ell}(Q^2)},
\end{align}
where the in-medium bubble diagram leads to
\begin{align}
\Pi^{* \ell\ell}_{\beta} (Q^2) &= \frac{3\,Q^2}{\pi^2} \int_0^1\!\! dx 
\int_{1/\Lambda_{\rm UV}^2}^{1/\Lambda_{\rm IR}^2} \frac{d\tau}{\tau}\
x\left(1-x\right)\, e^{-\tau\left[M_q^{* 2} + x\left(1-x\right)Q^2\right]}.
\end{align}

In the limit of $Q^2 \to \infty$, these form factors reduce to the elementary quark charges as expected. 
For small $Q^2$, they are similar to the expectations of the vector meson dominance hypothesis, 
where the $u$ and $d$ quarks are dressed by $\rho$ and $\omega$ mesons. 
The denominator in Eqs.~\eqref{eq:f1U} has the same pole structure of the Bethe-Salpeter equation 
in the $\rho$ or $\omega$ channels. 
Therefore, the dressed $u$ and $d$ quark form factors have poles at $Q^2 = -m_\rho^2$ and $Q^2 = -m_\omega^2$.

The final expression for the in-medium pion form factor with the dressed quark-photon vertex is then written as 
\begin{align}
\label{eq:fullPion}
F^{*}_{\pi^{+}}(Q^2) &= [F^{*}_{1U}(Q^2) - F^{*}_{1D}(Q^2)]\,
f^{*  \ell \ell}_\pi (Q^2).
\end{align} 
More detailed discussions on the pion form factor can be found, for example, in Refs.~\cite{HBCT18,HCT16,CBT14}.

\subsection{Results} \label{results}

\begin{figure}[t]
\centering\includegraphics[width=0.95\columnwidth]{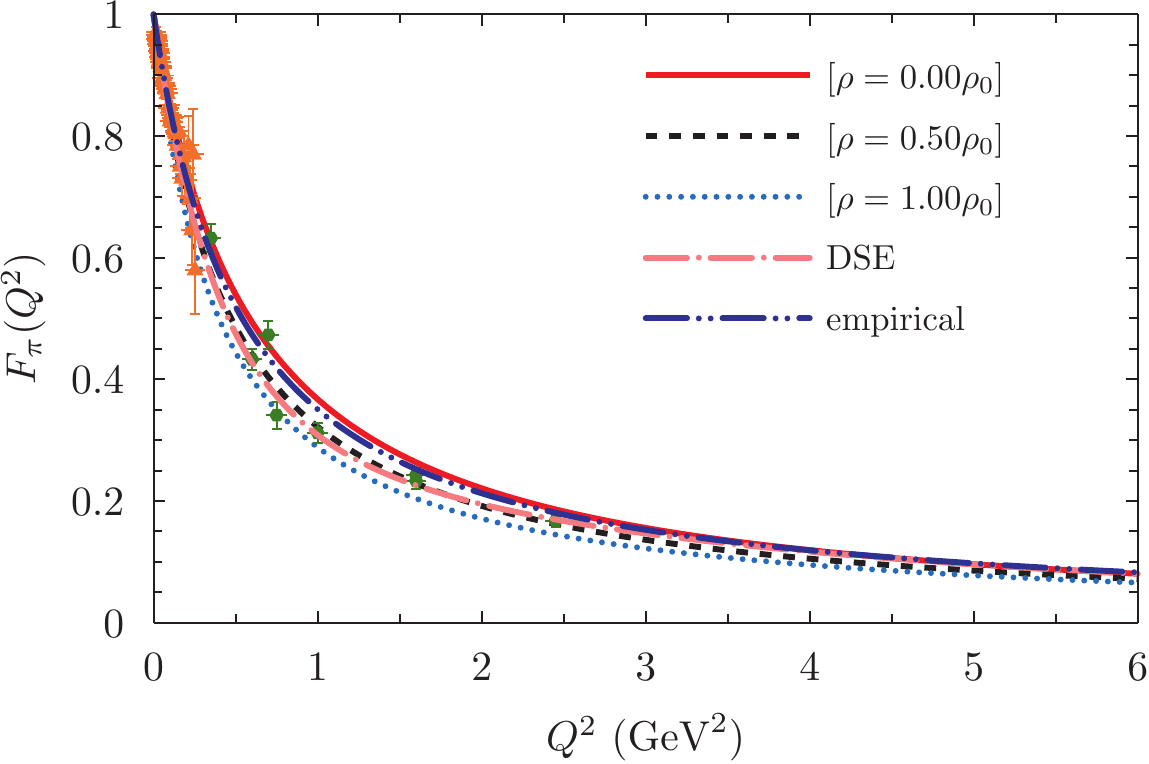}
\caption{\label{fig10} 
The electromagnetic form factor of positively charged pion for different nuclear matter densities.
The solid, dashed, and dotted lines are for $\rho_B^{} / \rho_0^{} = 0.0$, $0.5$, and $1.0$, respectively. 
The solid line is to be compared with the DSE predictions of Ref.~\cite{CCRST13},
the empirical parameterization of Ref.~\cite{ABBB84},
and the experimental data from Refs.~\cite{NA7-86,ABBB84,JLAB_Fpi-08}.}
\end{figure}

\begin{figure}[t]
\centering\includegraphics[width=0.95\columnwidth]{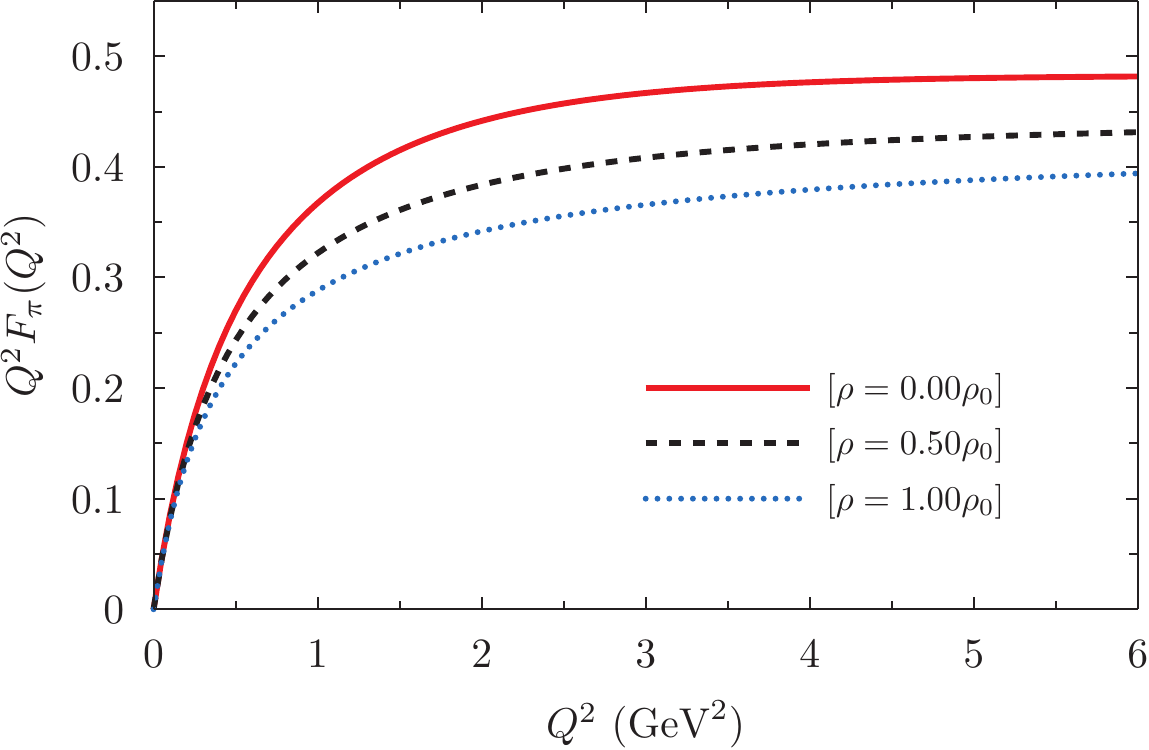}
\caption{\label{fig11} 
Results for $Q^2 F_{\pi}^{*} (Q^2)$ for various nuclear matter densities.
}
\end{figure}

With the dressed quark-photon vertex, our numerical results for the space-like electromagnetic form factors of the positively 
charged in-medium pion are shown in Figs.~\ref{fig10} and~\ref{fig11}. 
In the present calculation, we assume $m_\pi^{*} \sim m_\pi$ up to $\rho \simeq 1.25 \, \rho_0$ as verified in 
the previous section.
In Fig.~\ref{fig10}, our results are compared with the experimental data for free pion form factor~\cite{NA7-86,ABBB84,JLAB_Fpi-08}, 
the Dyson-Schwinger equation (DSE) result of Ref.~\cite{CCRST13}, and the empirical parameterization of Ref.~\cite{ABBB84}.

Figure~\ref{fig10} shows that the $Q^2$ dependence of the pion electromagnetic form factor  
for several nuclear matter densities, namely, $\rho_B^{} /\rho_0^{} = 0.0$, $0.5$, and $1.0$. 
Our results for the free-space pion form factor (solid line) are in good agreement with the data 
and the empirical monopole function form of $F_\pi (Q^2) = [1 +Q^2/\Lambda_\pi^2]^{-1}$~\cite{ABBB84}
with $\Lambda_\pi^2 =$ 0.54\,$\textrm{GeV}^2$. 
Our results show that the in-medium pion electromagnetic form factor is suppressed with increasing density.

Shown in Fig.~\ref{fig11} are the same results as in Fig.~\ref{fig10} but for $Q^2 F_\pi (Q^2)$.
The medium effects on the suppression of the pion form factor are clearly seen and it is reduced by
about 20\% at normal nuclear density, which would be large enough to be extracted empirically.

With the obtained electromagnetic form factors of the pion, the root-mean-square charge radius of the pion and 
its dependence on density can also be explored.
The charge radius is calculated through
\begin{equation}
\langle r_\pi \rangle = 
\sqrt{\left. -6 \frac{d F_\pi (Q^2)}{d Q^2}\right |_{Q^2 =0}}.
\end{equation}
Results for the in-medium pion charge radius for various nuclear densities are listed in Table~\ref{tab:model6}. 
The obtained charge radius of the pion in vacuum is in good agreement with the empirical data~\cite{PDG16}. 
Our calculations show that the charge radius of the in-medium pion increases with density and at normal nuclear density
it is predicted to be larger than the free-space value by about 20\%.

\begin{table}[t!]
\caption{Charge radius $r_\pi^{}$ of the positively charged pion in symmetric nuclear matter in units of fm.
}
\label{tab:model6}
\addtolength{\tabcolsep}{6.8pt}
\begin{tabular}{ccc} 
\hline \hline
$\rho_B^{} / \rho_0^{}$ & $r_\pi^{}$  &$r^{\rm expt.}$~\cite{PDG16} \\[0.2em] 
\hline
$0.00$ & $0.629$ &  $0.672 \pm 0.008$  \\ 
$0.25$ & $0.655$ &   \\
$0.50$ & $0.683$ &    \\
$0.75$ & $0.713$ &    \\
$1.00$ & $0.747$ &    \\
$1.25$ & $0.802$ &  
\\ \hline \hline
\end{tabular}
\end{table}

\section{Summary} \label{summary}

In the present work, the electroweak properties of the pion in symmetric nuclear matter 
have been explored in the framework of the NJL model.
The in-medium modifications of quark properties are estimated by the QMC model and are used as inputs of the NJL model calculations.
By combining the NJL and QMC models, we have investigated the density dependence of the quark condensates, 
the pion decay constant, the pion-quark coupling constant, and the pion mass.

We have found that the ratios of the in-medium to vacuum quark condensates, pion-quark-quark coupling constant, and
the pion decay constant decrease as the nuclear matter density increases. 
Our estimates on the ratios of $ \braket{ \bar{q} q }^{*} / \braket{ \bar{q} q }$, $g_{\pi q q}^{*} / g_{\pi q q}$, and 
$f_\pi^{*} / f_\pi$ are $0.87$, $0.77$, and $0.87$, respectively, at normal nuclear density.

Then the space-like electromagnetic form factor of the in-medium pion is calculated by considering 
the dressed quark-photon vertex and its density dependence.
By adopting the calculated in-medium pion-quark coupling constant, we estimated the pion electromagnetic form factor in medium.
The in-medium pion form factor is found to be suppressed compared with the form factor in vacuum as the nuclear 
matter density increases, while the charge radius of the pion is found to increase as density increases.
These effects are as large as 20\% of the vacuum values and would be tested by experiments at present and future facilities~\cite{CB99}.

\begin{acknowledgments}

K.T. thanks the Asia Pacific Center for Theoretical Physics (APCTP) for warm hospitality and excellent supports 
during his visit and stay. 
P.T.P.H. was supported by the Ministry of Science, ICT and Future Planning,
Gyeongsangbuk-do and Pohang City through the Young Scientist Training Asia-Pacific Economic Cooperation 
program of APCTP. 
The work of Y.O. was supported by the National Research Foundation of Korea under Grant 
Nos.\ NRF-2018R1D1A1B07048183 and NRF-2018R1A6A1A06024970.
The work of K.T. was supported by the Conselho Nacional de Desenvolvimento Cient\'{i}fico e Tecnol\'{o}gico - CNPq  
Grants, No.~400826/2014-3 and No.~308088/2015-8, and was also part of the projects, Instituto Nacional de Ci\^{e}ncia e 
Tecnologia - Nuclear Physics and Applications (INCT-FNA), Brazil, Process No. 464898/2014-5, and FAPESP Tem\'{a}tico, 
Brazil, Process No. 2017/05660-0.

\end{acknowledgments}

\end{document}